\documentclass[11pt]{elsarticle}

\usepackage{amsmath, amssymb, amsfonts}
\usepackage{graphicx}

\usepackage{xcolor,soul}%%%%%%%%%%%%%%%%%%%%%%%%%%%%%%%%%%%%%%%%%%%%%%%%%%%%%%%
\setstcolor{red}%%%%%%%%%%%%%%%%%%%%%%%%%%%%%%%%%%%%%%%%%%%%%%%%%%%%%%%%%%%%%%%
\journal{Chaos, Solitons & Fractals}

\begin{document}

\begin{frontmatter}

\title{Thermodynamic analysis of diverse percolation transitions}

\author[1]{Seonghyeon Moon}
\author[1,2]{Young Sul Cho\corref{cor1}}
\ead{yscho@jbnu.ac.kr}
\cortext[cor1]{Corresponding author}

\affiliation[1]{
    organization={Department of Physics, Jeonbuk National University},
    city={Jeonju},
    postcode={54896},
    country={Rep. of Korea}
}

\affiliation[2]{
    organization={Research Institute of Physics and Chemistry, Jeonbuk National University},
    city={Jeonju},
    postcode={54896},
    country={Rep. of Korea}
}

\begin{abstract}
This work extends the thermodynamic analysis of random bond percolation to explosive and hybrid percolation models.
We show that this thermodynamic analysis is well applicable to both explosive and hybrid percolation models by using the critical exponents $\alpha$ and $\delta$ obtained from scaling relations with previously measured values of $\beta$ and $\gamma$ within the error range. As a result, Rushbrooke inequality holds as an equality, $\alpha + 2\beta + \gamma = 2$, in both explosive and hybrid percolation models, where $\alpha > 0$ leads to the divergence of specific heats at the critical points.
Remarkably, entropy clearly reveals a continuous decrease even in a finite-sized explosive percolation model, unlike the order parameter. In contrast, entropy decreases discontinuously during a discontinuous transition in a hybrid percolation model, resembling the heat outflow during discontinuous transitions in thermal systems.
\end{abstract}


\begin{highlights}
% \item \cnew{3 to 5 bullet points, maximum 85 characters, including spaces, per bullet point}
\item The theoretical framework for the thermodynamic analysis of percolation is extended.
\item Thermodynamic scaling behaviors in diverse percolation transitions are confirmed.
\item Entropy decreases continuously during the explosive percolation transition (EPT).
\item Entropy decreases discontinuously during the hybrid percolation transition (HPT).
\item Specific heat diverges in both EPT and HPT, in contrast to random bond percolation.
\end{highlights}

\begin{keyword}
Explosive percolation transition \sep Hybrid percolation transition \sep Thermodynamics
\end{keyword}

\end{frontmatter}


\section{Introduction}
\label{sec:intro}

Percolation is a phenomenon where fluid penetrates through opposite boundaries of a system~\cite{stauffer, kim_percolation}, such as in sol-gel transitions~\cite{flory1} and metal--insulator transitions~\cite{con_insul, last:1971}.  
To understand the emergence of such phenomena, percolation models are used that describe the formation of clusters as bonds are occupied on a given lattice with $N$ nodes. The control parameter $p$ represents the fraction of occupied bonds, and a giant cluster of macroscopic size emerges at the critical point $p_c$. 
The evolution of the cluster size distribution $n_s$ is studied as a function of $p$, where $n_s$ represents the number of $s$-sized clusters over $N$.
In this paper, we examine three types of percolation models: (random) bond percolation, explosive percolation, and hybrid percolation.

In the bond percolation model, each bond on a given lattice is occupied randomly with the probability $p$.
Using the Fortuin--Kasteleyn formulation~\cite{KFformula}, the free energy of the Potts model on a given lattice is related to $G=\sum_s n_s(p)e^{-hs}+h$ of bond percolation on the same lattice, through the relation $p=1-e^{-K}$ where $K$ is the inverse temperature and $h$ is the magnetic field of the Potts model. Using this relation, percolation analogues of thermodynamic states were proposed in previous studies~\cite{KirkpatrickPRL1976, EssamJPhysC1971, StanleyJPhysA1978, FYWuJStatPhys1978}, as listed in Table~\ref{Table:thermodynamics}. The scaling behaviors of the thermodynamic states near the critical point of bond percolation were then studied in diverse lattices~\cite{KirkpatrickPRL1976, StanleyJPhysA1978, StanleyPRB1980}.

In this paper, we extend the previous results of bond percolation to the explosive and hybrid percolation models. Specifically, we use the product rule and half-restricted process to implement the explosive and hybrid percolation models, respectively, on a two-dimensional square lattice. In Sec.~\ref{sec:review}, we summarize the previous results of the thermodynamic analysis of bond percolation and confirm these results through simulation. In Sec.~\ref{sec:epmodel}, we demonstrate that the theoretical framework for bond percolation is also applicable to an explosive percolation model, which exhibits a continuous transition as shown through simulation. Then in Sec.~\ref{sec:hybridmodel} we modify the theoretical framework for the thermodynamic analysis of continuous transitions for application to hybrid percolation models and confirm its validity through simulation.
In Sec.~\ref{sec:conclusion}, we conclude by discussing the results. The Appendix provides a detailed description of the numerical methods used in this paper.

\begin{table*}
\centering
\caption{Percolation analogues of thermodynamic states as functions of two independent variables, $\varepsilon \equiv (p_c-p)/p_c$ and $h$. $G^{(1)}(\varepsilon, h)$ and $G^{(2)}(\varepsilon, h)$ denote the first- and second-order partial derivatives of $G(\varepsilon, h)$ with respect to $h$, respectively.}
\begin{tabular}{|c|c|}
\hline
               & Percolation analogue of thermodynamic states  \\
\hline
Free energy    &         $G(\varepsilon,h)=\sum_s n_s(p)e^{-hs}+h$       \\
\hline
Order parameter  &         $M(\varepsilon,h)\equiv G^{(1)}(\varepsilon,h)=1-\sum_s sn_s(p)e^{-hs}$    \\
\hline
Susceptibility &         $\chi(\varepsilon,h)\equiv G^{(2)}(\varepsilon,h)=\sum_s s^2n_s(p)e^{-hs}$    \\
\hline
Entropy        &         $S(\varepsilon,h)=-(1-p)\sum_s\frac{\partial n_s}{\partial p}e^{-hs}$    \\
\hline
Specific heat  &         $C_h(\varepsilon,h)=(1-p)\frac{\partial}{\partial p}\big[(1-p)\sum_s\frac{\partial n_s}{\partial p}e^{-hs}\big]$                                      \\
\hline
\end{tabular}
\label{Table:thermodynamics}
\end{table*}

\section{Thermodynamic analysis of bond percolation}
\label{sec:review}

In the bond percolation model, the scaling hypothesis of the free energy near the critical point $p_c$ is governed by two scaling powers, $a_{\varepsilon}, a_{h}$, as 
\begin{equation}
\widetilde{G}(\lambda^{a_{\varepsilon}}\varepsilon,  \lambda^{a_h}h) = \lambda \widetilde{G}(\varepsilon, h),
\label{eq:Gscaling}
\end{equation}
where $\widetilde{G}(\varepsilon, h)$ is the singular part of $G$.
Accordingly, the scaling hypothesis of $G^{(1)}$ and $G^{(2)}$ is given by
\begin{equation}
\lambda^{a_h}G^{(1)}(\lambda^{a_{\varepsilon}}\varepsilon, \lambda^{a_h}h)=\lambda G^{(1)}(\varepsilon, h),
\label{eq:G1scaling}
\end{equation}
\begin{equation}
\lambda^{2a_h}G^{(2)}(\lambda^{a_{\varepsilon}}\varepsilon, \lambda^{a_h}h)=\lambda G^{(2)}(\varepsilon, h),
\label{eq:G2scaling}
\end{equation}
where $G^{(1)}=\widetilde{G}^{(1)}$ and $G^{(2)}=\widetilde{G}^{(2)}$~\cite{StanleyJPhysA1978}.

For $h > 0$, the scaling behaviors of $G^{(1)}$ and $G^{(2)}$ near $h=0$ are related to the percolation critical exponent $\delta$. Specifically, the scaling behaviors are given by
\begin{equation}
G^{(1)}(\varepsilon, h)=h^{(1-a_h)/a_h}G^{(1)}(h^{-a_{\varepsilon}/a_h}\varepsilon, 1),
\label{eq:G1scalingh}
\end{equation}
\begin{equation}
G^{(2)}(\varepsilon, h)=h^{(1-2a_h)/a_h}G^{(2)}(h^{-a_{\varepsilon}/a_h}\varepsilon, 1),
\label{eq:G2scalingh}
\end{equation}
where the percolation critical exponent $\delta$ satisfies $1/\delta = (1-a_h)/a_h$.

For $h=0$, the scaling behaviors of $\widetilde{G}, G^{(1)}, G^{(2)}$ near $|\varepsilon|=0$ are related to percolation critical exponents $\alpha, \beta, \gamma$, respectively.  
At first, the scaling behavior of $\widetilde{G}$ is given by
\begin{align}
\widetilde{G}(\varepsilon, 0) &= 
  \begin{cases}
    \varepsilon^{1/a_{\varepsilon}}\widetilde{G}(1,0) & \text{~~for~~} \varepsilon > 0,\\
    (-\varepsilon)^{1/a_{\varepsilon}}\widetilde{G}(-1,0) & \text{~~for~~} \varepsilon \leq 0.
  \end{cases}
  \label{eq:Gscalingh0}
\end{align}
By considering the scaling behavior $\widetilde{G}(\varepsilon, 0) \sim |\varepsilon|^{2-\alpha}$, $1/a_{\varepsilon}=2-\alpha$ is obtained. 
Next, the scaling behavior of $G^{(1)}(\varepsilon, 0)$ is given by
\begin{align}
G^{(1)}(\varepsilon, 0) &= 
  \begin{cases}
    0 & \text{~~for~~} \varepsilon > 0,\\
    (-\varepsilon)^{(1-a_h)/a_{\varepsilon}}G^{(1)}(-1,0) & \text{~~for~~} \varepsilon \leq 0.
  \end{cases}
  \label{eq:G1scalingh0}
\end{align}
In this case, by considering the scaling behavior $G^{(1)}(\varepsilon, 0) \equiv M(\varepsilon, 0) \sim (-\varepsilon)^{\beta}$ for $\varepsilon \leq 0$,
$(1-a_h)/a_{\varepsilon}=\beta$ is obtained.
We note that $M(\varepsilon, 0)$ represents the fraction of nodes that belong to the giant cluster, which serves as the conventional order parameter in percolation models.
Finally, the scaling behavior of $G^{(2)}(\varepsilon, 0)$ is given by
\begin{align}
G^{(2)}(\varepsilon, 0) &= 
  \begin{cases}
    \varepsilon^{-(2a_h-1)/a_{\varepsilon}}G^{(2)}(1,0) & \text{~~for~~} \varepsilon > 0,\\
    (-\varepsilon)^{-(2a_h-1)/a_{\varepsilon}}G^{(2)}(-1,0) & \text{~~for~~} \varepsilon < 0.
  \end{cases}
  \label{eq:G2scalingh0}
\end{align}
By considering the scaling behavior $G^{(2)}(\varepsilon, 0) \equiv \chi(\varepsilon, 0) \sim |\varepsilon|^{-\gamma}$, $(2a_h-1)/a_{\varepsilon}=\gamma$ is obtained.
We note that $\chi(\varepsilon, 0)$ is the second moment of the cluster size distribution, which serves as the susceptibility in percolation models.

The scaling relations derived from Eq.~(\ref{eq:Gscaling})
indicate that the values of each critical exponent $\alpha$ and $\gamma$ are identical in the subcritical ($\varepsilon > 0$) and supercritical ($\varepsilon < 0$) regions.
Moreover, the two exponents $\delta$ and $\alpha$ are related to the two independent critical exponents $\beta$ and $\gamma$
such that $1/\delta = \beta/(\beta+\gamma)$ and $2-\alpha = 2\beta+\gamma$. It is noteworthy that the Rushbrooke inequality holds as an equality, $\alpha + 2\beta + \gamma = 2$.

\begin{figure}
	\includegraphics[width=0.8\textwidth]{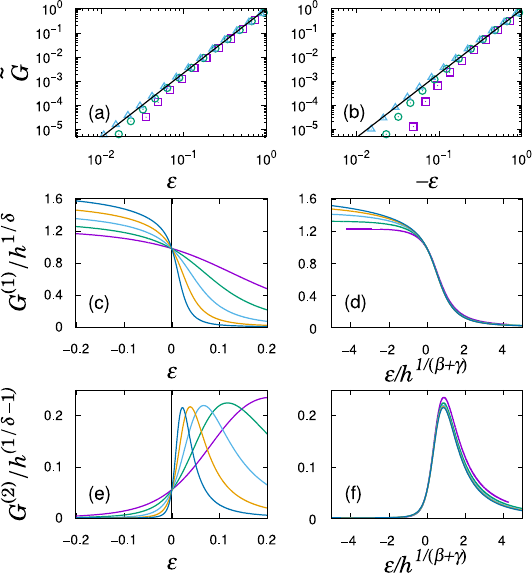} 
	\caption{Scaling behaviors of the free energy of bond percolation. (a,b) $\widetilde{G}(\varepsilon, 0)$ for $N=2^{16}, 2^{20}, 2^{24}$ from right to left. The slope of the solid lines is $8/3$. (c) $G^{(1)}(\varepsilon, h)/h^{1/\delta}$ for $h/10^{-4}=1, 4, 16, 64, 256$ in decreasing order of steepness. The vertical solid line indicates $\varepsilon=0$. (d) Data collapse of the results in (c) near $\varepsilon=0$. (e) $G^{(2)}(\varepsilon, h)/h^{1/\delta-1}$ for $h/10^{-4}=1, 4, 16, 64, 256$ in decreasing order of steepness, where the vertical solid line indicates $\varepsilon=0$. (f) Data collapse of the results in (e) near $\varepsilon=0$.}
	\label{Fig:bondPT_scaling}
\end{figure}

We confirm the scaling behaviors on the two-dimensional square lattice with $N$ nodes in Fig.~\ref{Fig:bondPT_scaling}, referring to previous results using the same model~\cite{StanleyJPhysA1978}. For further details on the numerical methods, refer to~\ref{app:bond}. Here, we use the theoretical values $p_c=0.5, \alpha = -2/3$, $\beta = 5/36$, $\gamma = 43/18$, and $\delta = 91/5$. The free energy for $h=0$ is expanded as $G(\varepsilon, 0)=a_0+a_1\varepsilon+a_2\varepsilon^2+C^{\pm}|\varepsilon|^{2-\alpha}$, retaining the dominant term of the singular part, where $a_0, a_1, a_2$ are constants independent of $\varepsilon$, and $C^{+}(C^{-})$ represents the coefficient for $\varepsilon<0 (\varepsilon > 0)$. We estimate $a_0, a_1, a_2$ using the relations $G(0,0)=a_0, G'(0, 0)=a_1, G''(0,0)=2a_2$, where $G'$ and $G''$ are the first- and second-order partial derivatives of $G$ with respect to $\varepsilon$. In Fig.~\ref{Fig:bondPT_scaling}(a) and (b), we confirm that $G(\varepsilon, 0)-a_0-a_1\varepsilon-a_2\varepsilon^2 \sim |\varepsilon|^{8/3}$ with the values of $a_0, a_1, a_2$ obtained by numerically computing the first and second derivatives of $G$ for each $N$. The estimated values in the thermodynamic limit are $a_0=0.1 \pm 0.003, a_1=0.5 \pm 0.003, a_2=1.35 \pm 0.05$, and $C^{\pm}=-1 \pm 0.3$.

In Fig.~\ref{Fig:bondPT_scaling}(c), $G^{(1)}(\varepsilon, h)/h^{1/\delta}$ for different $h$ cross at $\varepsilon=0$. In Fig.~\ref{Fig:bondPT_scaling}(d), $G^{(1)}(\varepsilon/h^{1/(\beta+\gamma)}, 1)/h^{1/\delta}$ for different $h$ collapse onto a single curve near $\varepsilon=0$. These behaviors in Fig.~\ref{Fig:bondPT_scaling}(c) and (d) confirm Eq.~(\ref{eq:G1scalingh}) with $1/\delta = (1-a_h)/a_h$ and $a_{\varepsilon}/a_h=1/(\beta+\gamma)$.
Then as shown in Fig.~\ref{Fig:bondPT_scaling}(e), $G^{(2)}(\varepsilon, h)/h^{(1/\delta-1)}$ for different $h$ cross at $\varepsilon=0$, while in Fig.~\ref{Fig:bondPT_scaling}(f), $G^{(2)}(\varepsilon/h^{1/(\beta+\gamma)}, 1)/h^{(1/\delta-1)}$ for different $h$ collapse onto a single curve near $\varepsilon=0$. As above, the behaviors in Fig.~\ref{Fig:bondPT_scaling}(e) and (f) confirm Eq.~(\ref{eq:G2scalingh}) with $1/\delta - 1 = (1-2a_h)/a_h$ and $a_{\varepsilon}/a_h=1/(\beta+\gamma)$.

\begin{figure}
	\includegraphics[width=0.8\textwidth]{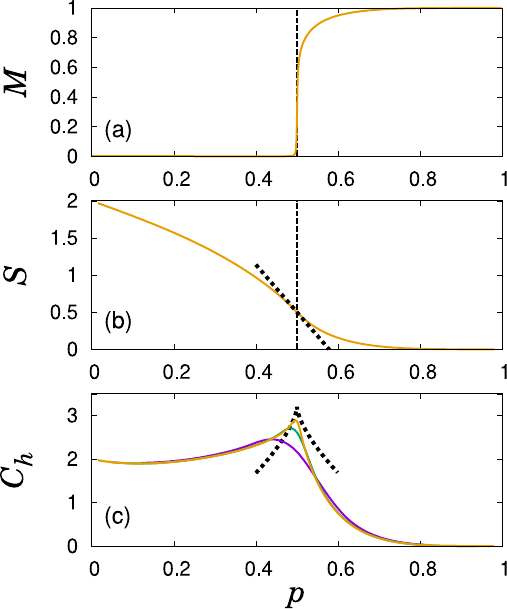} 
	\caption{Thermodynamic states of bond percolation. (a) $M(\varepsilon, 0)$ with $N=2^{24}$. The dashed line indicates $p_c$. (b) $S(\varepsilon, 0)$ with $N=2^{24}$. The vertical dashed line indicates $p_c$, and the dotted line depicts the critical behavior near $p_c$. (c) $C_h(\varepsilon, 0)$ with $N=2^{16}, 2^{20}, 2^{24}$ from bottom to top. The dotted line illustrates the critical behavior near $p_c$.}
	\label{Fig:bondPT_Pinf_S_C}
\end{figure}

We now discuss the derivation of the percolation analogues of the thermodynamic states $M, \chi, S$, and $C_h$ presented in Table.~\ref{Table:thermodynamics}.
First, it is natural that $M(\varepsilon, h)=1-\sum_ssn_se^{-hs}=G^{(1)}(\varepsilon,h)$ and $\chi=G^{(2)}(\varepsilon,h)$ for the percolation analogue of free energy $G(\varepsilon,h)$ following thermodynamic relations. Here, we discuss the role of $h$ using the bond occupation probability $p_0=1-e^{-h}$ between each node and an external node, a so-called ghost site~\cite{GriffithsGhostfield, StanleyGhostfield}. 

We write the order parameter as $M = 1 - \sum_s sn_s (1-p_0)^s$. If each node of an $s$-sized cluster is connected to the ghost site with probability $p_0$, the cluster merges into an infinite cluster through the ghost site with the probability $1-(1-p_0)^s$. Therefore, $M$ represents the probability that each node belongs to an infinite cluster when a magnetic field is applied, and it increases with $p_0$ such that the magnetic field $h$ enhances the global ordering by increasing $p_0$.
We note that $M$ and $\chi$ are written as $M=(1-p_0)\partial{G}/\partial{p_0}$ and $\chi = (1-p_0)\partial{M}/\partial{p_0}$ using the occupation probability $p_0$.
Accordingly, the percolation analogues of entropy and specific heat are given by $S=-(1-p)\partial{G}/\partial{p}$ and $C_h=-(1-p)\partial{S}/\partial{p}$, where $1-p$ is the probability that each pair of neighbors on the lattice is disconnected due to disordering, which increases as the temperature increases~\cite{StanleyJPhysA1978, KirkpatrickPRL1976}.

In Fig.~\ref{Fig:bondPT_Pinf_S_C}, we focus on the thermodynamic states for $h=0$, which are determined by the occupied bonds on the lattice without considering the ghost site.
In Fig.~\ref{Fig:bondPT_Pinf_S_C}(a), $M(\varepsilon, 0)$ becomes $M(\varepsilon, 0)>0$ continuously for $\varepsilon < 0$. 
The scaling behaviors of $S$ and $C_h$ near $\varepsilon=0$ for $h=0$ can be derived from the scaling behaviors of $G$ using the relations $S(\varepsilon, 0)=[\varepsilon+(1-p_c)/p_c]G'(\varepsilon, 0)$ and $C_h(\varepsilon, 0)=[\varepsilon+(1-p_c)/p_c]S'(\varepsilon, 0)$. As a result,
\begin{equation}
S(\varepsilon, 0)=a_1 + (a_1+2a_2)\varepsilon
\end{equation}
near $\varepsilon=0$, and
\begin{align}
C_h(\varepsilon, 0) &= 
  \begin{cases}
    a_1+2a_2+\frac{40}{9}C^-\varepsilon^{2/3} \text{~~for~~} \varepsilon > 0 \\\\
    a_1+2a_2+\frac{40}{9}C^+(-\varepsilon)^{2/3} \text{~~for~~} \varepsilon \leq 0
  \end{cases}
  \label{eq:bondC}
\end{align}
near $\varepsilon=0$. In Fig.~\ref{Fig:bondPT_Pinf_S_C}(b) and (c), the simulation results of $S$ and $C_h$ approach these theoretical values with the estimated values of $a_1, a_2$, and $C^{\pm}$ as $N$ increases.

\section{Results}
\subsection{Thermodynamic analysis of explosive percolation transitions}
\label{sec:epmodel}

An explosive percolation model suppresses the growth of large clusters, causing the giant cluster size to increase rapidly at $p_c$~\cite{Achlioptas:2009, explosive_phenomena, souza_nphy}. We are interested in an explosive percolation model in which the growth of large clusters is locally suppressed and a continuous transition is exhibited~\cite{hklee, dacosta_prl, dacosta_pre, grassberger, riordan, smoh:2016}. In this case, the thermodynamic analysis of the bond percolation model in the previous section may be applicable to the explosive percolation model, as both exhibit continuous transitions.

In this section, we implement a representative explosive percolation model, known as the product rule, on a two-dimensional square lattice. This model has been studied in several papers~\cite{ziff_prl, ziff_lattice:2010, filippo_pre:2010, Youjin_PRR2024, Jan_gap2}. Although the measured values from previous works are consistent within the error range, we specifically adopt the values of $p_c, \beta, \gamma$ from the latest study~\cite{Youjin_PRR2024}, which determined the critical exponents using the finite-size scaling method in accordance with standard finite-size scaling theory~\cite{Youjin_PRL2023}. Using these adopted critical exponent values, we confirm that the scaling hypothesis for the thermodynamic states of bond percolation is also applicable to the explosive percolation model. For further details on the numerical methods, refer to~\ref{app:ep}. We note that the main difference here from bond percolation is that $\alpha>0$, causing $C_h$ to diverge at $p_c$.

\begin{figure}
	\includegraphics[width=0.8\textwidth]{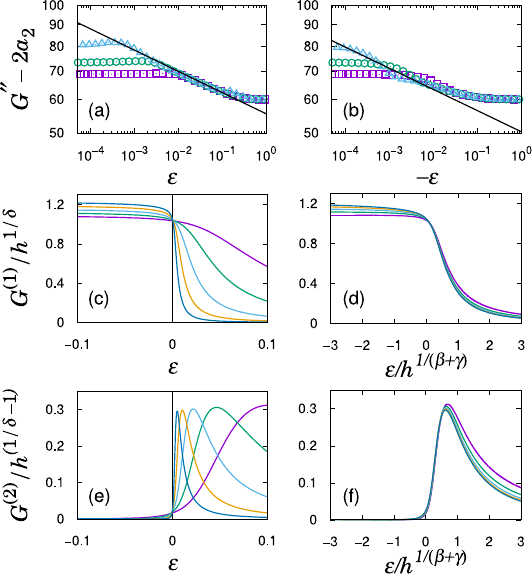} 
	\caption{Scaling behaviors of the free energy of explosive percolation. (a,b) $G''(\varepsilon,0)-2a_2$ with $a_2=-30$ for $N=2^{16}, 2^{20}, 2^{24}$ from bottom to top. The slope of the solid lines is $-0.05$. In (c--f), $1/\delta=0.022$ and $\beta+\gamma=1.907$ are used. (c) $G^{(1)}(\varepsilon, h)/h^{1/\delta}$ for $h/10^{-4}=1, 4, 16, 64, 256$ in decreasing order of steepness. The vertical solid line indicates $\varepsilon=0$. (d) Data collapse of the results in (c) near $\varepsilon=0$. (e) $G^{(2)}(\varepsilon, h)/h^{1/\delta-1}$ for $h/10^{-4}=1, 4, 16, 64, 256$ from left to right, with the vertical solid line indicating $\varepsilon=0$. (f) Data collapse of the results in (e) near $\varepsilon=0$.}
	\label{Fig:PR_scaling}
\end{figure}

Using the measured values from~\cite{Youjin_PRR2024} and the scaling relations between the critical exponents, we obtain $p_c=0.5266 \pm 0.0001$, $1/\delta = 0.022 \pm 0.001$, and $\beta+\gamma=1.907 \pm 0.001$. Then by using the two additional relations $1/\delta = \beta/(\beta+\gamma)$ and $2-\alpha=2\beta+\gamma$, we obtain $\beta=0.042\pm0.001, \gamma=1.865\pm 0.001$, and $\alpha = 0.051 \pm 0.001$.

Here, we find that $\alpha>0$, unlike in bond percolation. Therefore, the free energy for $h=0$ is expanded as
$G(\varepsilon, 0)=a_0+a_1\varepsilon+C^{\pm}|\varepsilon|^{2-\alpha}+a_2\varepsilon^2$ up to $\mathcal{O}(\varepsilon^2)$ in ascending order of exponents, where $a_0, a_1, a_2$ are constants independent of $\varepsilon$ and $C^{+}(C^{-})$ represents the coefficient for $\varepsilon<0 ( > 0)$.
In this case $a_1=0.2685 \pm 0.001$ as estimated by $G'(0,0)=a_1$.
Since $\alpha$ has a small positive value, the term $2a_2$ in $G''(\varepsilon, 0)$ should be taken into account to fit the simulation data. Accordingly, $G''(\varepsilon, 0)=C^{\pm}(2-\alpha)(1-\alpha)|\varepsilon|^{-\alpha}+2a_2$ up to $\mathcal{O}(1)$.
In Fig.~\ref{Fig:PR_scaling}(a) and (b), we confirm that $G''(\varepsilon, 0)-2a_2 \sim |\varepsilon|^{-\alpha}$ with the estimated values $a_2 = -30 \pm 5$, $C^{+}=27 \pm 5$, and $C^{-} = 30 \pm 5$.
The scaling behaviors in Eqs.~(\ref{eq:G1scalingh}) and (\ref{eq:G2scalingh}) are confirmed in Fig.~\ref{Fig:PR_scaling}(c,d) and (e,f), respectively, using the relations $(1-a_h)/a_{\varepsilon}=\beta$ and $(2a_h-1)/a_{\varepsilon}=\gamma$.

\begin{figure}
	\includegraphics[width=0.8\textwidth]{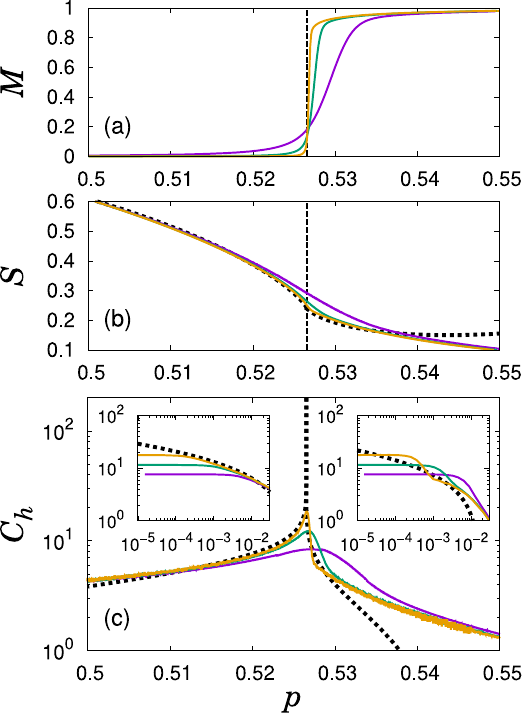} 
	\caption{Thermodynamic states of explosive percolation. (a) $M(\varepsilon, 0)$ with $N=2^{16}, 2^{20}, 2^{24}$ in increasing order of steepness. The dashed line indicates $p_c$. (b) $S(\varepsilon, 0)$ with $N=2^{16}, 2^{20}, 2^{24}$ from top to bottom. The vertical dashed line indicates $p_c$, and the dotted line depicts the critical behavior near $p_c$. (c) $C_h(\varepsilon, 0)$ with $N=2^{16}, 2^{20}, 2^{24}$ from bottom to top. The dotted line illustrates the critical behavior near $p_c$. Insets: $C_h(\varepsilon, 0)$ of the main panel vs. $p_c-p$ (left) and $p-p_c$ (right).}
	\label{Fig:PR_Pinf_S_C}
\end{figure}

In Fig.~\ref{Fig:PR_Pinf_S_C}, we focus on the thermodynamic states for $h = 0$. It can be seen in Fig.~\ref{Fig:PR_Pinf_S_C}(a) that $M(\varepsilon, 0)$ increases more sharply near $\varepsilon=0$ as $N$ increases, even though it exhibits a continuous transition at $\varepsilon=0$ in the thermodynamic limit $N \rightarrow \infty$.
The scaling behaviors of $S(\varepsilon, 0)$ and $C_h(\varepsilon, 0)$ near $\varepsilon=0$ are given by
\begin{align}
S(\varepsilon, 0) &= 
  \begin{cases}
    \frac{a_1(1-p_c)}{p_c} + C^-\frac{(1-p_c)(2-\alpha)}{p_c}\varepsilon^{1-\alpha} \\ +\Big[\frac{2a_2(1-p_c)}{p_c}+a_1\Big]\varepsilon \text{~~~~for~~} \varepsilon > 0, \\\\
   \frac{a_1(1-p_c)}{p_c} - C^+\frac{(1-p_c)(2-\alpha)}{p_c}(-\varepsilon)^{1-\alpha} \\ +\Big[\frac{2a_2(1-p_c)}{p_c}+a_1\Big]\varepsilon \text{~~~~for~~} \varepsilon \leq 0,
  \end{cases}
  \label{eq:PRS}
\end{align}
up to $\mathcal{O}(\varepsilon)$, and by
\begin{align}
C_h(\varepsilon, 0) &= 
  \begin{cases}
    C^-\frac{(1-p_c)^2(2-\alpha)(1-\alpha)}{p_c^2}\varepsilon^{-\alpha} \\ +
    \frac{(1-p_c)}{p_c}\Big[a_1+\frac{2a_2(1-p_c)}{p_c}\Big] \text{~~~~for~~} \varepsilon > 0, \\\\
   C^+\frac{(1-p_c)^2(2-\alpha)(1-\alpha)}{p_c^2}(-\varepsilon)^{-\alpha}
   \\ +\frac{(1-p_c)}{p_c}\Big[a_1+\frac{2a_2(1-p_c)}{p_c}\Big]  \text{~~~~for~~} \varepsilon \leq 0,
  \end{cases}
  \label{eq:PRC}
\end{align}
up to $\mathcal{O}(1)$. In Fig.~\ref{Fig:PR_Pinf_S_C}(b) and (c), the simulation results of $S$ and $C_h$ approach Eqs.~(\ref{eq:PRS}) and (\ref{eq:PRC}) with $p_c=0.5266$, $a_1=0.2685$, $a_2=-30$, $C^-=29.95$, and $C^+=27.15$ as $N$ increases.

We note that $M(\varepsilon, 0)$ increases so rapidly in Fig.~\ref{Fig:PR_Pinf_S_C}(a) that the model could initially be thought to exhibit a discontinuous transition. However, $S(\varepsilon, 0)$ clearly displays a continuous shape during the transition, as shown in Fig.~\ref{Fig:PR_Pinf_S_C}(b) and Eq.~(\ref{eq:PRS}), indicating that heat does not flow out as $\varepsilon$ falls below 0, and that the transition is indeed continuous. Additionally, 
$C_h(\varepsilon, 0)$ diverges as $C_h(\varepsilon, 0) \sim |\varepsilon|^{-\alpha}$, as shown in Eq.~(\ref{eq:PRC}),
in contrast to the behavior of $C_h(\varepsilon, 0)$ in bond percolation, which converges to a finite value at $\varepsilon = 0$ as discussed in the previous section.

\subsection{Thermodynamic analysis of hybrid percolation transitions}
\label{sec:hybridmodel}

We now consider a hybrid percolation model, in which the growth of large clusters is globally suppressed. As a result, both a discontinuous and a continuous transition occur at the same $p_c$~\cite{choprl:2016, park_hybrid}. Specifically, $M(\varepsilon, 0) - M_0 \sim (-\varepsilon)^{\beta}$ for $\varepsilon \leq 0$ and $M(\varepsilon, 0)=0$ for $\varepsilon > 0$, where $M_0>0$ is the discontinuity of $M(\varepsilon, 0)$ at $\varepsilon = 0$. 
Therefore, the free energy $G$ includes an additional non-singular term, $M_0 h$ for $\varepsilon \leq 0$, which reveals asymmetric behaviors before and after the threshold. 

Beginning with Eq.~(\ref{eq:Gscaling}), we use the modified relation $\widetilde{G}^{(1)}(\varepsilon, h)=G^{(1)}(\varepsilon, h)-M_0$.
Then Eq.~(\ref{eq:G1scaling}) should be modified as
\begin{equation}
\lambda[G^{(1)}(\varepsilon, h)-M_0] = \lambda^{a_h}[G^{(1)}(\lambda^{a_{\varepsilon}}\varepsilon, \lambda^{a_h}h)-M_0]
\label{eq:G1scalinghybrid}
\end{equation}
for $\varepsilon \leq 0$ by replacing $G^{(1)}(\varepsilon, h)$ with $G^{(1)}(\varepsilon, h)-M_0$.
From Eq.~(\ref{eq:G1scalinghybrid}), the scaling behaviors of $G^{(1)}$ for $h > 0$ and $h=0$ are given by
\begin{equation}
G^{(1)}(\varepsilon, h)-M_0=h^{(1-a_h)/a_h}[G^{(1)}(h^{-a_{\varepsilon}/a_h}\varepsilon, 1)-M_0],
\label{eq:G1scalinghhybrid}
\end{equation}
and
\begin{equation}
G^{(1)}(\varepsilon, 0)-M_0=(-\varepsilon)^{(1-a_h)/a_{\varepsilon}}[G^{(1)}(-1,0)-M_0],
\label{eq:G1scalingh0hybrid}
\end{equation}
respectively, with the scaling relations $1/\delta = (1-a_h)/a_h$ and $\beta = (1-a_h)/a_{\varepsilon}$ for $\varepsilon \leq 0$.
We note that Eq.~(\ref{eq:G1scalingh0hybrid}) clearly reproduces the hybrid transition $M(\varepsilon, 0)-M_0 \sim (-\varepsilon)^{\beta}$.

Also from Eq.~(\ref{eq:G1scalinghybrid}), the scaling behaviors of $G^{(2)}$ for $h > 0$ and $h=0$ are given by
\begin{equation}
G^{(2)}(\varepsilon, h)=h^{(1-2a_h)/a_h}G^{(2)}(h^{-a_{\varepsilon}/a_h}\varepsilon, 1),
\label{eq:G2scalinghybrid}
\end{equation}
and
\begin{equation}
G^{(2)}(\varepsilon, 0)=(-\varepsilon)^{-(2a_h-1)/a_{\varepsilon}}G^{(2)}(-1, 0),
\end{equation}
respectively, with the scaling relation 
$(2a_h-1)/a_{\varepsilon}=\gamma$ for $\varepsilon \leq 0$.

Based on the simulation results shown below, we assume that the expansion of $G(\varepsilon, 0)$ is given by
\begin{align}
G(\varepsilon, 0) &= 
  \begin{cases}
    a_0 + a_1^-\varepsilon+C^-\varepsilon^{2-\alpha} &\text{for~~} \varepsilon > 0, \\\\
   a_0 + a_1^+\varepsilon+C^+(-\varepsilon)^{2-\alpha} &\text{for~~} \varepsilon \leq 0,
  \end{cases}
  \label{eq:Gexpansionhybrid}
\end{align}
up to $\mathcal{O}(\varepsilon^{2-\alpha})$ with $\alpha>0$, where $a_0$ is a constant independent of $\varepsilon$, and $a_1^+$ $(a_1^-)$ and $C^{+}(C^{-})$ represent the coefficients for $\varepsilon<0 ( > 0)$. Eq.~(\ref{eq:Gexpansionhybrid}) implies that the scaling hypothesis of $\widetilde{G}(\varepsilon, 0) \sim |\varepsilon|^{2-\alpha}$ of the hybrid percolation transition would also be given by Eq.~(\ref{eq:Gscalingh0}) with the scaling relation $1/a_{\varepsilon}=2-\alpha$.

The scaling relations in Eqs.~(\ref{eq:G1scalinghybrid}--\ref{eq:Gexpansionhybrid}) indicate that the two exponents $\delta$ and $\alpha$ are related to the two independent critical exponents $\beta$ and $\gamma$ such that $1/\delta=\beta/(\beta+\gamma)$ and $2-\alpha = 2\beta + \gamma$. It is noteworthy that the Rushbrooke inequality holds as an equality, $\alpha + 2\beta + \gamma = 2$, even in the hybrid percolation transition.

Now we aim to provide a heuristic argument regarding why the coefficients $a_1^- = \lim_{\varepsilon \to 0^+} G'(\varepsilon, 0)$ and $a_1^+ = \lim_{\varepsilon \to 0^-} G'(\varepsilon, 0)$ are different in this case, in contrast to in continuous transitions such as bond percolation and explosive percolation. $G(\varepsilon, 0) = \sum_s n_s$ represents the number of clusters per site, which decreases by $1/N$ when an occupied bond connects two distinct clusters but remains unchanged when an occupied bond is within a cluster. Therefore, $G'(\varepsilon, 0)$ is proportional to the probability that an occupied bond connects two distinct clusters, as $\varepsilon$ decreases linearly with the number of occupied bonds.

We let $f^-$ and $f^+$ denote the probabilities that an occupied bond connects two distinct clusters in the limits $\varepsilon \rightarrow 0^+$ and $\varepsilon \rightarrow 0^-$, respectively. Thus $a_1^- \propto f^-$ and $a_1^+ \propto f^+$. During the transition $|\varepsilon| \ll 1$, we approximate the hybrid percolation model using bond percolation.
Then, $f^-$ is equivalent to the fraction of unoccupied bonds between two distinct clusters in the limit $\varepsilon \rightarrow 0^+$. 
When a giant cluster of size $M_0$ emerges in the limit $\varepsilon \rightarrow 0^-$, it is assumed that the probabilities of neighboring nodes belonging to the giant cluster are uncorrelated and each equal to $M_0$. Under this assumption, the relation between $f^+$ and $f^-$ is given by $f^+= (1-M_0^2)f^-$.
As a result, $a_1^+/a_1^-=f^+/f^-=1-M_0^2$, and $a_1^--a_1^+=M_0^2a_1^->0$.

The scaling behavior of $S(\varepsilon, 0)$ is given by
\begin{align}
S(\varepsilon, 0) &= 
  \begin{cases}
    \frac{a_1^-(1-p_c)}{p_c} + C^-\frac{(1-p_c)(2-\alpha)}{p_c}\varepsilon^{1-\alpha} &\text{~~~~for~~} \varepsilon > 0, \\\\
   \frac{a_1^+(1-p_c)}{p_c} - C^+\frac{(1-p_c)(2-\alpha)}{p_c}(-\varepsilon)^{1-\alpha} &\text{~~~~for~~} \varepsilon \leq 0,
  \end{cases}
  \label{eq:HRS}
\end{align}
up to $\mathcal{O}(|\varepsilon|^{1-\alpha})$. As expected, $S(\varepsilon, 0)$ decreases discontinuously at $\varepsilon=0$, indicating that heat flows out as $\varepsilon$ falls below $0$, and that the transition is indeed discontinuous. We note that the discontinuity of $G'(\varepsilon, 0)$ at $\varepsilon=0$, given by $a_1^- -a_1^+$, induces the discontinuity of $S(\varepsilon, 0)$.

The scaling behavior of $C_h(\varepsilon, 0)$ is given by 
\begin{align}
C_h(\varepsilon, 0) &= 
  \begin{cases}
    C^-\frac{(1-p_c)^2(2-\alpha)(1-\alpha)}{p_c^2}\varepsilon^{-\alpha} &\text{for~~} \varepsilon > 0, \\\\
   C^+\frac{(1-p_c)^2(2-\alpha)(1-\alpha)}{p_c^2}(-\varepsilon)^{-\alpha} &\text{for~~} \varepsilon \leq 0,
  \end{cases}
  \label{eq:HRC}
\end{align}
up to $\mathcal{O}(|\varepsilon|^{-\alpha})$.

\begin{figure}
	\includegraphics[width=0.8\textwidth]{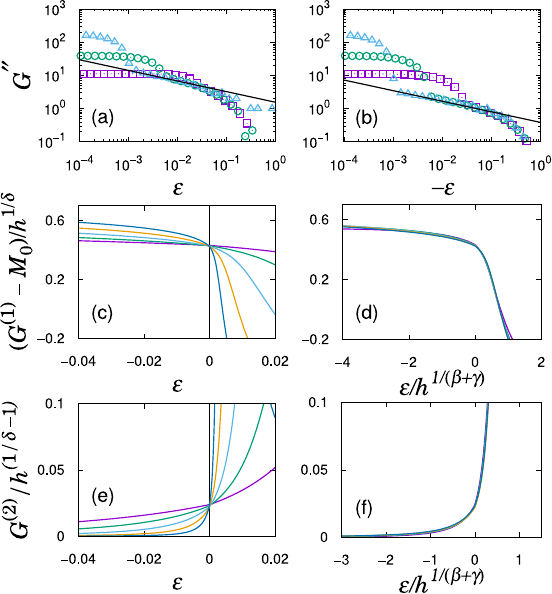} 
	\caption{Scaling behaviors of the free energy of hybrid percolation. (a,b) $G''(\varepsilon,0)$ for $N=2^{16}, 2^{20}, 2^{24}$ from bottom to top. The slope of the solid lines is $-0.32$. (c) $(G^{(1)}(\varepsilon, h)-M_0)/h^{1/\delta}$ for $h/10^{-4}=1, 4, 16, 64, 256$ from left to right. The vertical solid line indicates $\varepsilon=0$. (d) Data collapse of the results in (c) near $\varepsilon=0$. (e) $G^{(2)}(\varepsilon, h)/h^{1/\delta-1}$ for $h/10^{-4}=1, 4, 16, 64, 256$ from left to right, with the vertical solid line indicating $\varepsilon=0$. (f) Data collapse of the results in (e) near $\varepsilon=0$.}
	\label{Fig:HR_scaling}
\end{figure}

\begin{figure}
	\includegraphics[width=0.8\textwidth]{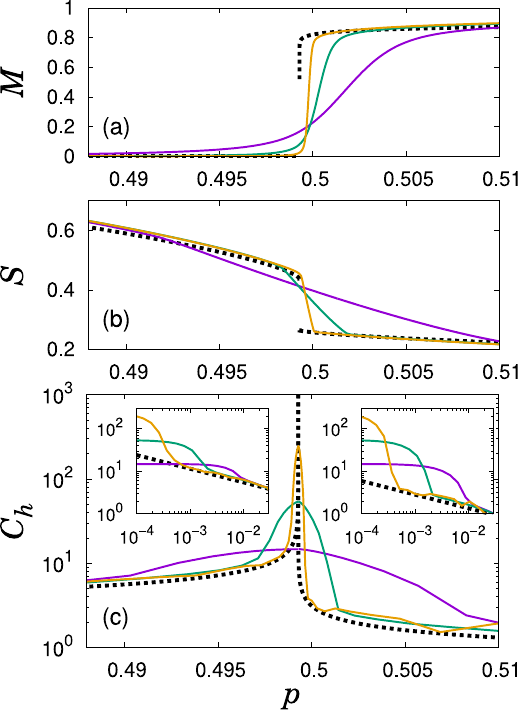} 
	\caption{Thermodynamic states of hybrid percolation. (a) $M(\varepsilon, 0)$ with $N=2^{16}, 2^{20}, 2^{24}$ in increasing order of steepness. The dotted line represents the estimated shape in the thermodynamic limit $N \rightarrow \infty$. (b) $S(\varepsilon, 0)$ with $N=2^{16}, 2^{20}, 2^{24}$ in increasing order of steepness. The dotted line represents the estimated shape in the thermodynamic limit $N \rightarrow \infty$. (c) $C_h(\varepsilon, 0)$ with $N=2^{16}, 2^{20}, 2^{24}$ from bottom to top. The dotted line illustrates the critical behavior near $p_c$. Insets: Specific heat of the main panel vs. $p_c-p$ (left) and $p-p_c$ (right).}
	\label{Fig:HR_Pinf_S_C}
\end{figure}

The scaling behavior of the hybrid percolation transition is confirmed using a hybrid percolation model~\cite{Half_restricted}, specifically the half-restricted process on a two-dimensional square lattice. We adopt the measured values of $M_0, \beta$, and $\gamma$ from a previous study that used the same model~\cite{PREhybrid2D} and show that the scaling behaviors are valid with the relations $1/\delta=\beta/(\beta+\gamma)$ and $\alpha=2-2\beta-\gamma$ within the error range of $\beta, \gamma$. For further details on the numerical methods, refer to~\ref{app:hybrid}.

The half-restricted process incorporates an external parameter $0 < g \leq 1$.
In~\cite{PREhybrid2D}, the reported values for $g=0.5$ are $M_0=0.55 \pm 0.03$, $\beta = 0.06 \pm 0.01$, and $\gamma = 1.56 \pm 0.15$, yielding $1/(\beta+\gamma) = 0.62 \pm 0.07$.
Accordingly, the estimated values $1/\delta = 0.04 \pm 0.01$ and $\alpha = 0.32 \pm 0.05$ are obtained from the scaling relations. Here, we find that $\alpha > 0$, in contrast to bond percolation.

In Fig.~\ref{Fig:HR_scaling}(a) we check that $G''(\varepsilon) \approx C^-(2-\alpha)(1-\alpha)\varepsilon^{-\alpha}$ with $C^-=1.36$ and $\alpha=0.32$ for $\varepsilon > 0$. Then in Fig.~\ref{Fig:HR_scaling}(b) we check that $G''(\varepsilon) \approx C^+(2-\alpha)(1-\alpha)(-\varepsilon)^{-\alpha}$ with $C^+=0.33$ and $\alpha=0.32$ for $\varepsilon < 0$.
The scaling behaviors in Eqs.~(\ref{eq:G1scalinghybrid}) and (\ref{eq:G2scalinghybrid}) are confirmed in Fig.~\ref{Fig:HR_scaling}(c,d) and (e,f), respectively, using
$M_0=0.55, 1/\delta = 0.05$, and $1/(\beta+\gamma)=0.57$, all of which fall within the error range.

The thermodynamic states for $h=0$ with the estimated $p_c = 0.4993 \pm 0.001$ are plotted in Fig.~\ref{Fig:HR_Pinf_S_C}. Here, $p_c$ slightly differs from the value in~\cite{PREhybrid2D} because we used a free boundary condition instead of a periodic boundary condition, in line with the other percolation models discussed in the previous sections. In Fig.~\ref{Fig:HR_Pinf_S_C}(a), $M(\varepsilon, 0)$ increases more sharply near $\varepsilon=0$ as $N$ increases. As a result, it exhibits a hybrid transition in the thermodynamic limit $N \rightarrow \infty$, following the dotted curve $M(\varepsilon, 0)-M_0 \sim (-\varepsilon)^{\beta}$
for $\varepsilon \leq 0$ with $\beta = 0.06$.
In Fig.~\ref{Fig:HR_Pinf_S_C}(b) and (c), the simulation results of $S$ and $C_h$ approach Eqs.~(\ref{eq:HRS}) and (\ref{eq:HRC}) with $p_c = 0.4993, a_1^- = 0.4365, a_1^+ = 0.2635, C^-=1.36, C^+=0.33$, and $\alpha = 0.32$ as $N$ increases.

We estimate $a_1^+ = 0.26 \pm 0.01$ and $a_1^- = 0.44 \pm 0.01$ such that the ratio $a_1^+/a_1^-=0.59 \pm 0.04$. This ratio is smaller than the value expected from the heuristic argument above, $1-M_0^2 = 0.70 \pm 0.04$. We assumed in the heuristic argument that the probabilities of neighboring nodes belonging to the giant cluster are uncorrelated and each equal to $M_0$; however, the probabilities of neighboring nodes belonging to the giant cluster are actually positively correlated. As a result, more than $M_0^2$ fraction of the unoccupied bonds between distinct clusters in the limit $\varepsilon \to 0^+$ would belong to the giant cluster in the limit $\varepsilon \to 0^-$. This implies that $f^+ < (1-M_0^2)f^-$ and thus $a_1^+/a_1^-=f^+/f^-<1-M_0^2$. We note that the discontinuity of $a_1^--a_1^+$ is still ensured by the condition $a_1^--a_1^+>a_1^-M_0^2>0$.

Finally, we interpret the effect of the global suppression rule in the hybrid percolation model from a thermodynamic perspective. Occupation of
links in a percolation model can be interpreted as a process of decreasing external temperature. In bond percolation without any suppression, the system remains in thermal equilibrium with the environment throughout the process. In contrast, the global suppression rule in the hybrid percolation model can be interpreted as a mechanism that restricts heat flow out of the system, thereby increasing the temperature gap between the system and the environment during the process. At the threshold, this temperature gap becomes sufficiently large to render the suppression mechanism ineffective, allowing the accumulated heat to be released.

\section{Conclusion}
\label{sec:conclusion}

In previous studies, the information entropy~\cite{Shannon} of individual clusters~\cite{EPentropyHassan1, EPentropyHassan2} and the cluster size distribution~\cite{EPentropyVieira, EPentropyCho} were examined as a function of $p$ in both bond percolation and explosive percolation models, and the specific heat of information entropy was also studied. The behavior of information entropy in those studies differs from the entropy behavior presented in this work. For example, the previously reported critical exponent $\alpha$ for specific heat differs from the value obtained in this work, and the information entropy of the cluster size distribution was reported to reach a maximum near the critical point, whereas the entropy in this study monotonically decreases as $p$ increases.  
It should be discussed why the results exhibit these different properties, even though both the information entropy of the previous studies and the entropy used here measure the uncertainty of the system.

In this paper, we extended the theoretical framework for the thermodynamic analysis of continuous transitions to include hybrid percolation transitions. We confirmed that the theoretical framework for continuous transitions is effective even in an explosive percolation model, and that our newly proposed framework for hybrid transitions applies well to the hybrid percolation model. The main difference between the two transitions is that, in a continuous transition, the giant cluster size does not jump at the threshold, resulting in symmetrical criticality before and after the threshold. In contrast, in a hybrid transition, there is a jump in the giant cluster size, leading to asymmetric criticality before and after the threshold~\cite{choi_hybrid}. Here, the scaling behaviors of the order parameter and susceptibility were analyzed only in the supercritical region ($\varepsilon \leq 0$), which directly corresponds to the continuous transition of the order parameter after the jump. The scaling behaviors in the subcritical region ($\varepsilon > 0$) have not yet been addressed. Therefore, the theoretical framework for thermodynamic analysis needs further exploration to cover both subcritical and supercritical regions of various hybrid transitions, including percolation transitions of $k$-core~\cite{kcore}, bootstrap~\cite{Bootstrap, BootstrapDorogovtsev}, and interdependent networks~\cite{InterdependentHavlin, InterdependentDorogovtsev}.

\appendix

\section{Numerical analyses for bond percolation}  \label{app:bond}

In the bond percolation analysis, we use a two-dimensional square lattice with $N$ nodes and $2\sqrt{N}(\sqrt{N}-1)$ bonds, applying free boundary conditions. We begin with all the bonds unoccupied at $p=0$. At each time step, $p \rightarrow p + 1/[2\sqrt{N}(\sqrt{N}-1)]$, we select one unoccupied bond randomly and occupy it. 
For a given value of $N$, parameters $G(\varepsilon,h), M(\varepsilon,h)$, and $\chi(\varepsilon,h)$ in Table~\ref{Table:thermodynamics} can be computed throughout the entire process $0 \leq p \leq 1$ for a single independent realization.
For a fixed value of $N$, we perform more than $10^5$ independent realizations and compute the averages of $G(\varepsilon, h), M(\varepsilon, h)$, and $\chi(\varepsilon, h)$ over these realizations.

$S(\varepsilon, h) = [\varepsilon + (1-p_c)/p_c]G'(\varepsilon, h)$ and $C_h(\varepsilon, h)=[\varepsilon + (1-p_c)/p_c]G'(\varepsilon, h)+[\varepsilon + (1-p_c)/p_c]^2G''(\varepsilon, h)$ are obtained through numerical differentiation of the averaged $G(\varepsilon, h)$.
Specifically, we use the formula $G'(\varepsilon, h)=[G(\varepsilon+\Delta \varepsilon, h)-G(\varepsilon-\Delta \varepsilon, h)]/(2\Delta \varepsilon)$ and
$G''(\varepsilon, h)=[G(\varepsilon+\Delta \varepsilon, h)+G(\varepsilon-\Delta \varepsilon, h)-2G(\varepsilon, h)]/(\Delta \varepsilon)^2$. 
To obtain $S$ and $C_h$ in Fig.~\ref{Fig:bondPT_Pinf_S_C}(b) and (c), we use $\Delta \varepsilon \propto N^{-3/8}$ derived from the relation $(\Delta \varepsilon)^{-\nu} \sim N^{1/2}$ with the correlation length exponent $\nu = 4/3$. Additionally, we use $\Delta \varepsilon \propto N^{-3/8}$ to obtain $G'(0, 0)$ and $G''(0, 0)$ to estimate $a_1$ and $a_2$.

\section{Numerical analyses for explosive percolation} \label{app:ep}
In the explosive percolation analysis, we use a two-dimensional square lattice with $N$ nodes and $2\sqrt{N}(\sqrt{N}-1)$ bonds, applying free boundary conditions. We begin with all the bonds unoccupied at $p=0$. At each time step, $p \rightarrow p + 1/[2\sqrt{N}(\sqrt{N}-1)]$, we select two unoccupied bonds randomly, among which the one with the smaller product of the connecting cluster sizes is occupied. For a given value of $N$, parameters $G(\varepsilon, h), M(\varepsilon, h)$, and $\chi(\varepsilon, h)$ in Table~\ref{Table:thermodynamics} can be computed throughout the entire process $0 \leq p \leq 1$ for a single independent realization. We use $p_c = 0.5266$, as measured in~\cite{Youjin_PRR2024}, for the relation $\varepsilon = (p_c-p)/p_c$.

$G^{(i)}, M^{(i)}$, and $\chi^{(i)}$ denote the values of $G, M$, and $\chi$ for each independent realization indexed by $i$, while $\varepsilon_c^{(i)} = (p_c-p_c^{(i)})/p_c$ is the peak point of $\chi^{(i)}(\varepsilon, 0)$. The averaged value $G(\varepsilon, h)$ is obtained by shifting the peak point of each $G^{(i)}$ to $\varepsilon = 0$ and then averaging it, yielding $G(\varepsilon, h)=\langle G^{(i)}(\varepsilon + \varepsilon_c^{(i)}, h)\rangle_i$, where $\langle \cdot \rangle_i$ denotes the average over $i$. As introduced in previous studies~\cite{Youjin_PRL2023, Youjin_PRR2024}, for the explosive percolation model, the average of an observable must be taken in this way to eliminate the effects of fluctuations in $p_c^{(i)}$ and to clearly obtain the scaling exponent. Similarly, the averages of $M$ and $\chi$ are obtained by $M(\varepsilon, h)=\langle M^{(i)}(\varepsilon+\varepsilon_c^{(i)}, h)\rangle_i$ and $\chi(\varepsilon, h)=\langle \chi^{(i)}(\varepsilon+\varepsilon_c^{(i)}, h)\rangle_i$.

$S(\varepsilon, h) = [\varepsilon + (1-p_c)/p_c]G'(\varepsilon, h)$ and $C_h(\varepsilon, h)=[\varepsilon + (1-p_c)/p_c]G'(\varepsilon, h)+[\varepsilon + (1-p_c)/p_c]^2G''(\varepsilon, h)$ are obtained through numerical differentiation of the averaged $G(\varepsilon, h)$.
Specifically, we use the formula $G'(\varepsilon, h)=[G(\varepsilon+\Delta \varepsilon, h)-G(\varepsilon-\Delta \varepsilon, h)]/(2\Delta \varepsilon)$ and
$G''(\varepsilon, h)=[G(\varepsilon+\Delta \varepsilon, h)+G(\varepsilon-\Delta \varepsilon, h)-2G(\varepsilon, h)]/(\Delta \varepsilon)^2$. 
To obtain $S$ and $C_h$ in Fig.~\ref{Fig:PR_Pinf_S_C}(b) and (c), we use $\Delta \varepsilon \propto N^{-0.5129}$ derived from the relation $(\Delta \varepsilon)^{-\nu} \sim N^{1/2}$ with the correlation length exponent $\nu = 1/1.0258$. Additionally, we use $\Delta \varepsilon \propto N^{-0.5129}$ to obtain $G''(\varepsilon, h)$ in Fig.~\ref{Fig:PR_scaling}(a) and (b).

\section{Numerical analyses for hybrid percolation} \label{app:hybrid}

In the hybrid percolation analysis, we use a two-dimensional square lattice with $N$ nodes and $2\sqrt{N}(\sqrt{N}-1)$ bonds, applying free boundary conditions. Here, an external parameter $0<g\leq 1$ is given. We begin with all the bonds unoccupied at $p=0$. At each time step, $p \rightarrow p + 1/[2\sqrt{N}(\sqrt{N}-1)]$, an unoccupied bond is occupied following steps (i--iii).
\begin{itemize}
    \item[(i)] A set of nodes $R(t)$ belonging to the $k$ smallest clusters is determined such that $\sum_{\ell=1}^{k-1}s_{\ell} < \lfloor gN \rfloor \leq \sum_{\ell=1}^k s_{\ell}$, where $s_{\ell}$ is the size of the $\ell$-th smallest cluster.
    \item[(ii)] An unoccupied bond is selected at random.
    \item[(iii)] If neither node of the selected bond belongs to $R(t)$, return to step (ii). Otherwise, occupy the unoccupied bond. 
\end{itemize}
For a given value of $N$, parameters $G(\varepsilon,h), M(\varepsilon,h)$, and $\chi(\varepsilon,h)$ in Table~\ref{Table:thermodynamics} can be computed throughout the entire process $0 \leq p \leq 1$ for a single independent realization. We use $p_c=0.4993$, as estimated in Sec.~\ref{sec:hybridmodel}, for the relation $\varepsilon = (p_c-p)/p_c$.

\begin{figure}
	\includegraphics[width=0.8\textwidth]{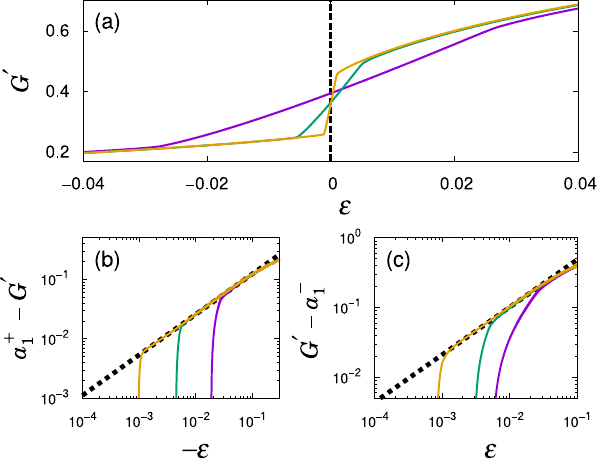} 
	\caption{Estimate of $a_1^+$ and $a_1^-$ in hybrid percolation. (a) $G'(\varepsilon, 0)$ for $N=2^{16}, 2^{20}, 2^{24}$ in increasing order of steepness. The vertical dashed line indicates $\varepsilon=0$. 
    (b) $a_1^+-G'(\varepsilon, 0)$ with $a_1^+ = 0.2635$ for $N=2^{16}, 2^{20}, 2^{24}$ from right to left, and
    (c) $G'(\varepsilon, 0)-a_1^-$ with $a_1^-=0.4365$ for $N=2^{16}, 2^{20}, 2^{24}$ from right to left. The slope of the dotted lines is $0.68$.}
	\label{Fig:HR_G_a1pa1m}
\end{figure}

$G^{(i)}, M^{(i)}$, and $\chi^{(i)}$ denote the values of $G, M$, and $\chi$ for each independent realization indexed by $i$, while $\varepsilon_c^{(i)} = (p_c-p_c^{(i)})/p_c$ is the peak point of $\chi^{(i)}(\varepsilon, 0)$.
To eliminate the effects of fluctuations in $p_c^{(i)}$ and clearly measure the critical exponents, the averages of $G$, $M$, and $\chi$ are calculated in the same manner as in the explosive percolation model. Specifically, $G(\varepsilon, h)=\langle G^{(i)}(\varepsilon+\varepsilon_c^{(i)}, h)\rangle_i$, $M(\varepsilon, h)=\langle M^{(i)}(\varepsilon+\varepsilon_c^{(i)}, h)\rangle_i$, and $\chi(\varepsilon, h)=\langle \chi^{(i)}(\varepsilon+\varepsilon_c^{(i)}, h)\rangle_i$, where $\langle \cdot \rangle_i$ denotes the average over $i$.

$S(\varepsilon, h) = [\varepsilon + (1-p_c)/p_c]G'(\varepsilon, h)$ and $C_h(\varepsilon, h)=[\varepsilon + (1-p_c)/p_c]G'(\varepsilon, h)+[\varepsilon + (1-p_c)/p_c]^2G''(\varepsilon, h)$ are obtained through numerical differentiation of the averaged $G(\varepsilon, h)$.
Specifically, we use the formula $G'(\varepsilon, h)=[G(\varepsilon+\Delta \varepsilon, h)-G(\varepsilon-\Delta \varepsilon, h)]/(2\Delta \varepsilon)$ and
$G''(\varepsilon, h)=[G(\varepsilon+\Delta \varepsilon, h)+G(\varepsilon-\Delta \varepsilon, h)-2G(\varepsilon, h)]/(\Delta \varepsilon)^2$. 
To obtain $S$ and $C_h$ in Fig.~\ref{Fig:HR_Pinf_S_C}(b) and (c), we use $\Delta \varepsilon \propto N^{-0.5945}$ derived from the relation $(\Delta \varepsilon)^{-\nu} \sim N^{1/2}$ with the correlation length exponent $\nu = 0.841$~\cite{choi_hybrid}. Additionally, we use $\Delta \varepsilon \propto N^{-0.5945}$ to obtain $G''(\varepsilon, h)$ in Fig.~\ref{Fig:HR_scaling}(a) and (b).

To estimate $a_1^+$ and $a_1^-$, we use the properties that $G'(\varepsilon, 0) - a_1^- \propto \varepsilon^{1-\alpha}$ for $\varepsilon > 0$ and $a_1^+  - G'(\varepsilon, 0) \propto (-\varepsilon)^{1-\alpha}$ for $\varepsilon \leq 0$, which follow from Eq.~(\ref{eq:Gexpansionhybrid}).
In Fig.~\ref{Fig:HR_G_a1pa1m}, we find that the power-law regime $G'(\varepsilon, 0)-a_1^-$ is the longest when $a_1^- \approx 0.4365$ and $\alpha \approx 0.32$, and that the power-law regime $a_1^+ - G'(\varepsilon, 0)$ is the longest when $a_1^+ \approx 0.2635$ and $\alpha \approx 0.32$. In the range of $\alpha = 0.32 \pm 0.05$, the estimated values are $a_1^- = 0.44 \pm 0.01$ and $a_1^+ = 0.26 \pm 0.01$.

\section*{Acknowledgments}
This work was supported by a National Research Foundation (NRF) of Korea grant, No. 2020R1F1A1061326.

\bibliographystyle{elsarticle-num.bst}
\bibliography{references.bib}

\end{document}